\documentclass[prl,preprint,showpacs,superscriptaddress]{revtex4}
\usepackage{graphicx}
\usepackage{dcolumn}

\begin{document}

\title{Steering effect on the shape of islands for homoepitaxial growth of
  Cu on Cu(001) }
\author{Jikeun Seo}
\affiliation{Division of General Education, Chodang University, 
  Muan 534-701, Republic of Korea }
\author{S.-M. Kwon}
\affiliation{Department of Physics, Sook-Myung Women's University,
  Seoul 140-742, Republic of Korea}
\author{H.-Y. Kim}
\affiliation{Department of Physics, Sook-Myung Women's University,
  Seoul 140-742, Republic of Korea}
\author{J.-S. Kim}
\affiliation{Department of Physics, Sook-Myung Women's University,
  Seoul 140-742, Republic of Korea}

\date{\today}

\begin{abstract}
The steering effect on the growth of islands is investigated by 
combining molecular dynamics (MD) and kinetic Monte Carlo (KMC) 
simulations. Dynamics of depositing atoms and kinetics of atoms on a 
substrate are realized by MD and KMC, respectively. The reported 
experimental results on the asymmetric island growth [van Dijken
{\it et al.}, Phys. Rev. Lett. {\bf 82}, 4038 (1999).] is well
reproduced. A salient phenomenon, the reversal of the asymmetry,
is found as the
island size increases, and  attributed to the asymmetric flux on the 
lower terrace of island.
\end{abstract}

\maketitle

 The growth of thin film is an essential step for many modern
technologies and scientific investigation, and a lot of efforts
have been made to understand and tailor the growth process. There 
have been many studies on the effect of energetic\cite{Bauer} and
kinetic\cite{Reviews} variables to the growth of thin film, while minor
attention has been paid to the role of dynamic variables such as
deposition conditions. Recently, van Dijken et al.\cite{Dijken}
reported the growth of rectangular Cu islands on square-symmetric Cu(001)
when the deposition was made at grazing incidence angle.
 Furthermore, they 
controlled the shape of Co island on Cu(001) by varying the deposition
angle and successfully manipulated its magnetic 
anisotropy\cite{Dijken2}. Their work clearly reveals the importance of
dynamic variable and expand the adjustable parameters for the 
growth of thin film.\par

  An atomistic picture for the asymmetric island growth was also proposed by
 van Dijken et al.\cite{Dijken}; Incident atoms see a modified
 potential by  pre-existing island
 and as a result the incoming flux is focused on the upper terrace 
 of the island near the front edge, while depleted
 on the lower terrace near the rear edge of the island. Such steering effect 
 is argued to result in the shortened
 edge length along the deposition direction, $x$-axis, and leaves the edge
 perpendicular to the deposition direction, $y$-axis, relatively
 longer. The model, however, is based on a qualitative argument
 without any detailed kinetic description on island growth from the
 inhomogeneous flux distribution, and the steering effect is considered
 only in the incident plane rather than in the three dimensional space
 as it should be. Independently, Zhong et al.\cite{Zhong} proposed that the
 enhanced corner crossing diffusion at the front edge due to the
 transient mobility of the deposited atom increased the growth speed
 along the {\it y}-direction relative to that along the $x$-direction.
 However, the very existence and the role of the transient mobility still
 remains controversial\cite{Sanders}. Steering effect has also been
 studied in the roughening in amorphous thin film growth by
 MD\cite{Landman} and non-linear stochastic equation\cite{Raible}.
 For the growth of a crystalline
 film, only recently Montalenti and Voter\cite{Montalenti} reported the
 steering induced instability of Ag film on Ag(001), but the study was
 restricted to the roughening of the film on a
 substrate at 0 K  and to the deposition at normal incidence.\par

   The purpose of the present study is to investigate the steering effect
 on the growth of asymmetric islands by performing a 
 realistic simulation combining KMC and MD simulation; MD simulation is
 executed to calculate the trajectory of depositing atoms in {\it three}
 dimensional space, only when a deposition event is selected in KMC.
 Before the next deposition event is selected, various diffusion
 events of adatoms are realized by KMC simulation\cite{Clancy}.
 In the present study, the asymmetric island shape observed in
 the experimental study\cite{Dijken} is well reproduced. Moreover,
 a salient phenomenon, the reversal of the asymmetry of island from
 elongation in the $y$-direction to the $x$-direction, is found as the coverage
 or the substrate temperature increases. The
 asymmetric shape of the island is attributed mainly to the asymmetric
 deposition flux on the lower terrace of island that in turn
 depends on its size.\par

 In MD simulation, Lennard-Jones potential in the form of
 $U(r) = 4 D [(\sigma /r)^{12}-(\sigma /r)^{6}]$ with $D=$0.4093 eV and
 $\sigma=$2.338 $\AA$ is used for the interaction between the incident
 atom and surface atoms\cite{Sanders}, and Verlet algorithm is adapted.
 The simulation box is composed of 6 layer high empty space in
 $z$-direction on an fcc(001) surface of 400$\times$400 atoms with
 periodic boundary condition in the plane. To reduce computational
 efforts, during deposition, the atoms on the substrate are frozen at each
 lattice site. (The surface lattice constant, $a_{0}=$2.56 $\AA$ and
 the interlayer spacing, $d=$1.805 $\AA$) The incident atom along [110]
 direction has initial kinetic energy of 0.15 eV determined from the
 melting temperature of Cu and follows the trajectory determined from
 the interaction potential starting at the initial height of $10 a_{0}$
 until the atom experiences the repulsive force by pre-existing atoms
 on the substrate. In the present simulation,
 {\it no transient mobility is taken into account}. \par

 For the diffusion processes, 12 different ones are taken into account 
 including those suggested by Furman et al.\cite{Furman}, and their diffusion
 barriers are adopted. Some of the most influential diffusion barriers are
 listed in Table 1.\par

 From the simulated morphology, the mound radius, which can directly be
 compared with the diffraction results\cite{Dijken}, is determined as 
 the first zero of the height-height correlation function
 $<h({\bf r})h(0)>-<h>^{2}$.
 To increase the statistical reliability, simulations 
 are performed 60 times under the same growth condition, and the mound
 radius is calculated over the whole set of simulated morphologies. \par

   In Fig. 1 we summarize the simulation results for the growth of Cu
 atoms on Cu(001) under the identical growth conditions to those of
 the previous experiment\cite{Dijken}. (We denote the mound radius
 along the $x$-axis by $L$ and that along the $y$-axis by $W$.) For
 the normal deposition, $W$ and $L$ are identical at all coverages,
 while  for the deposition at an angle, $80^o$ off the surface normal,
 the symmetry is broken and the difference between $W$ and $L$ becomes
 larger as the coverage increases. At 0.5 monolayer (ML), $W$ is larger
 than L by 5\% reproducing the previously reported result\cite{Dijken}.
 It is noteworthy that we reproduce the experimental
 result\cite{Dijken} without taking any transient mobility into account.
 The inset of Fig. 1 shows that for constant coverage of 0.5 ML,
 the aspect ratio, $W/L$, increases, as does the deposition angle.
 This result demonstrates that the asymmetry of the island shape is
 closely related to the reduced symmetry of the deposition geometry.\par

 The top figure in Fig. 2(a) reproduces the flux distribution along the
 $x$-axis through the center of a pre-existing island as reported in
 Ref. 3, which herein is called the front and rear edge({\it FR-edge})
 steering effect. Also found in the present study, as one may see in
 the following figures of Fig. 2(a), is the strong dependence of the
 flux distribution on the island size; as the island size becomes
 small, the portion of the island with enhanced flux is enlarged.
 Besides, in Fig. 2(b), an additional steering
 effect called herein the side edge({\it S-edge}) steering is observed;
 the trajectories of incident atoms are curved toward both side edges
 and enhanced flux is found along both side edges. As will be
 discussed later, both the dependence of deposition flux on the
 island size and the relative strength of {\it FR-edge} steering
 to {\it S-edge} steering have great impact on shaping the island.\par

 To have a comprehensive picture on the asymmetric island growth,
 a series of simulation is performed with various substrate 
 temperatures. Fig. 3 shows the dependence of $L$ and $W$ on
 coverage at various growth temperatures and the inset summarizes the
 temperature dependence of the aspect ratio at 0.5 ML. A stunning
 new phenomenon, the reversal of the asymmetry of island around
 230K, is found; $W>L$ above 230K as observed in Ref. 3, but
 $W<L$ below 230K. Such a reversal of the island shape has never
 been predicted before.\par

 We find that aforementioned, apparently diverse island growth can be
 understood by exploring the nature of the inhomogeneous deposition
 flux. The inhomogeneous flux is expected to manifest itself in two
 different ways; 1) asymmetric downhill current from the top of the
 island and 2) asymmetric deposition flux around the lower terrace of
 the island. As for the first one, despite the inhomogeneous deposition
 flux on the upper
 terrace of the island as shown in Fig. 2, the ratio of the downhill
 current per unit edge length over $x$-directional edge to that 
over $y$-directional edges is very
 close to 1.0 when we count the atoms coming over the {\it ES} barrier
 during growth simulation. This direction-independence of the downhill
 diffusion current occurs because the terrace diffusion on top of the
 island is more frequent by $\sim 10^2$ times than the downhill
 diffusion over {\it ES} barrier, and effectively homogenizes the
 inhomogeneous deposition flux
 on the island before atoms come down over the
 barrier. \par

  As for the second one, we investigate the flux distribution on the
 lower terrace of the island. In Fig. 4(a), shown is the average deposition
 flux on terraces within 3 $a_0$ from the side edges (S-terrace) and
 that from the front and rear edges (FR-terrace), respectively, as a
 function of the island size. The flux on the FR-terrace decreases
 monotonically, but that on the S-terrace shows a maximum as the
 island becomes larger. For the island with its edge length smaller than
 10 $a_0$, the flux on FR-terrace is larger than that on S-terrace, but
 the other way is true for the larger island.\par

  Aforementioned dependence of the deposition flux on the island size
 can be understood from the flux distribution shown in Figs. 2 and 5.
 For small islands(Fig. 2(a) middle and bottom), the perturbed potentials
 at front and rear edges interfere and result in the increased flux
 on the FR-terrace. As the island size increases, the flux near
 the rear edge becomes relatively small (Fig. 2(a) top) due to the
 uninterfered rear edge potential. Hence, the {\it average} flux on
 the FR-terrace monotonically decreases as the island size increases
 as found in Fig. 4(a).\par

  On the other hand, the average flux on the S-terrace is influenced
 by two counter-acting factors; firstly, the flux on the S-terrace
 increases with the island size, because the longer the edges the
 more atoms moving parallel to the side edge are attracted to the
 edge. However, the flux on S-terrace is not homogeneous along the
 side edge and another intricate variable is involved (Fig. 5). For small
 islands (Fig. 5 top and middle), the deposition flux is enhanced over
 the whole edge. As the island size increases, the relative portion of
 the side edge with pronounced flux enhancement gradually
 decreases(Fig. 5 bottom). This suggests that the {\it average} flux over
 the side edge will gradually decrease as the edge length increases.
 Hence, the competition between these two factors results in the maximum
 of the average flux at the island size of 17-20 $a_0$, explaining the
 dependence of the average flux on the island size observed in Fig. 4(a).
 The origin of the inhomogeneous enhancement of flux along the side
 edges on S-terraces can be found by the comparison of Figs. 2(a) and 5.
 The flux profile along the side edge(Fig. 5) shows
 quite a resemblance with that along the deposition direction
 through the center of the island (Fig. 2(a)).
 It suggests that the side edge flux is determined not solely by the {\it
 S-edge} steering but {\it cooperatively} by the {\it FR-edge}
 steering.\par

  The island-size dependence of the flux ratio as shown in Fig. 4(b)
 is quite similar to the temperature dependence of the aspect
 ratio of the islands as shown in the inset
 of Fig. 3. Keeping in mind that the mean island size
 monotonically increases with the substrate temperature for the same
 coverage, it suggests that the temperature dependence of the aspect
 ratio of the island may originate from the dependence of the flux
 ratio on the island size; At low substrate temperature where small
 islands form, the average flux on FR-terrace is larger than that on
 S-terrace and results in asymmetric island having $L>W$, and
 {\it vice versa}. We examine this idea further by plotting the aspect ratio
 of the islands as a function of the mean size of islands formed 
with varying growth conditions. 
As shown  in Fig. 4(b), even though
 the data are collected from uncorrelated simulations, there exists a strong
 correlation between the aspect ratio and the island size. Furthermore, it
 reproduces all the features found in the dependence of the flux ratio
 on the island size, especially the crossover (aspect ratio$=$1), maximum,
 and thereafter monotonic decrease of the ratio.
 (The island sizes where the crossover and the maximum in the
 aspect ratio occur are larger than the corresponding ones in the flux
 ratio. This is understood from the fact that the island size is
 determined by the integration of deposition flux.) It suggests that the most
 important factor determining the shape of the island is the deposition
 flux on the lower terrace of the island, which in turn is determined by
 the size of the island. If that conjecture is correct, the asymmetric
 island shape ($L>W$) should also be observed in the initial deposition
 stage and so does the reversal of the asymmetry with further growth
 unless the temperature is too low to form islands larger
 than the crossover size, because
 the island size is always small at the early stage of the island
 growth, irrespective of the substrate temperature.
 As indicated with $+$ signs in Fig. 3, the reversal of the
 asymmetry is indeed observed at similar island sizes, regardless of the
 substrate temperatures above 230K, reassuring
 that the island-size dependent flux ratio at the lower terrace of the
 island is the critical factor for shaping the island.\par

  In summary, the present simulation combining MD and KMC
 properly reproduces the previously reported rectangular island growth
 of Cu/Cu(001) under deposition at a grazing incidence angle, 
 and shows that the transient mobility is not a necessary
 condition for the asymmetric island growth. Instead, the
 asymmetry of the island shape is attributed mainly to the asymmetry
 of the deposition flux on the lower terrace of islands.
 Also found is the existence of the reversal of the aspect ratio of
 asymmetric islands, depending on the substrate temperature or the
 island size. This, so far, unanticipated phenomenon remains to be
 proven by further experimental study. It is also found important
 that, for a proper interpretation of the experimental results, the
 steering effect should be treated in full three dimension.

\begin{acknowledgments}
  This work is support by KISTEP(00-B-WB-07-A-01) (JK).
\end{acknowledgments}

\pagebreak

 \begin{table}
 \caption{\label{table1} Some of the diffusion barriers and diffusion
parameters used in simulation. Notations in the bracket 
[ ] are from Furman {\it et al.}\cite{Furman}.}
 \begin{ruledtabular}
 \begin{tabular}{cc}
 type of diffusion & diffusion barrier \\
 \tableline
 free jump [$E_{0}$] & 0.485 eV\\
 dimer lateral bond break [$E_{2}$] & 0.463 eV \\
 re-estabilishing NN bond [$E_{4}$] & 0.183 eV \\
 Erlich-Schw\"obel(ES) barrier($\Delta E_S$) & 0.1 eV\\
 \tableline
 jump frequency($\nu_{0}$) & $2.4 \times 10^{13} $ \\
 deposition rate ($F_{0}$) & 0.00416 ML/s \\
 \end{tabular}
 \end{ruledtabular}
 \end{table}

\pagebreak

\begin{figure}
\includegraphics[scale=0.8]{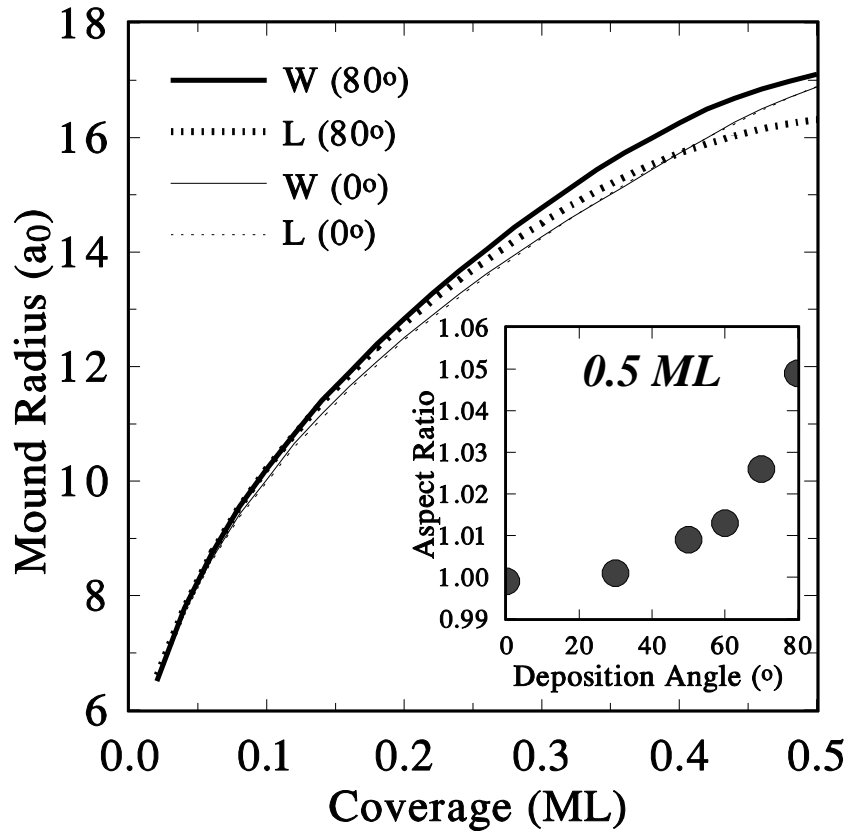}
\caption{\label{fig1} Mound radius along both x and y directions for
Cu islands grown on Cu(001) at 250 K. Inset : Aspect ratio of mound
radii, $W/L$, versus deposition angle (from the surface normal).}
\end{figure}

\begin{figure}
\includegraphics[scale=0.8]{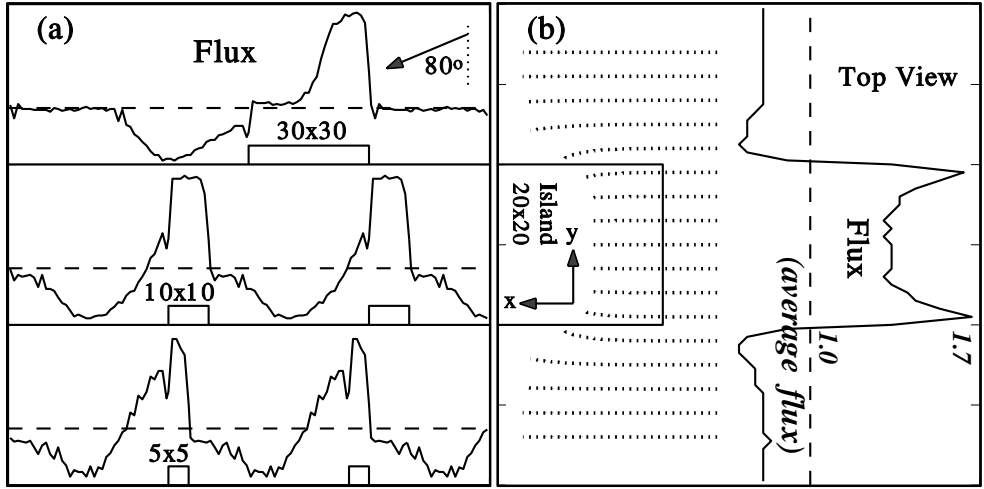}
\caption{\label{fig2} The deposition flux and atomic trajectory
calculated from MD simulation at deposition angle of 80 degree. 
(a) Deposition flux along x direction for islands of sizes,
30 $a_{0}$(top), 10 $a_{0}$(middle), and 5 $a_{0}$(bottom), respectively.
(b)Top view of the trajectory of incident atoms near a pre-existing
island, and flux distribution across the deposition direction, y
direction. Long dashed line represents the normalized flux. }
\end{figure}

\begin{figure}
\includegraphics[scale=0.8]{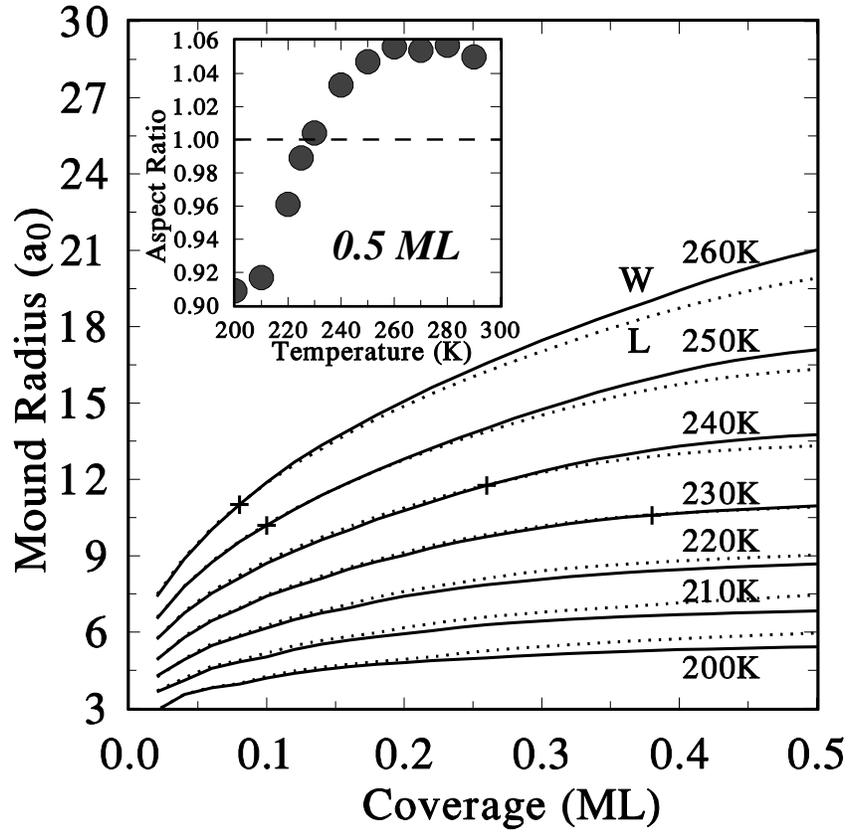}
\caption{\label{fig3} Temperature dependence of mound radius. Solid
curves represent mound radius($W$) along $y$-direction and dotted curves
 correspond to that ($L$) along $x$-direction. Inset : Aspect ratio 
($W/L$) for 0.5 ML growth. }
\end{figure}

\begin{figure}
\includegraphics[scale=0.8]{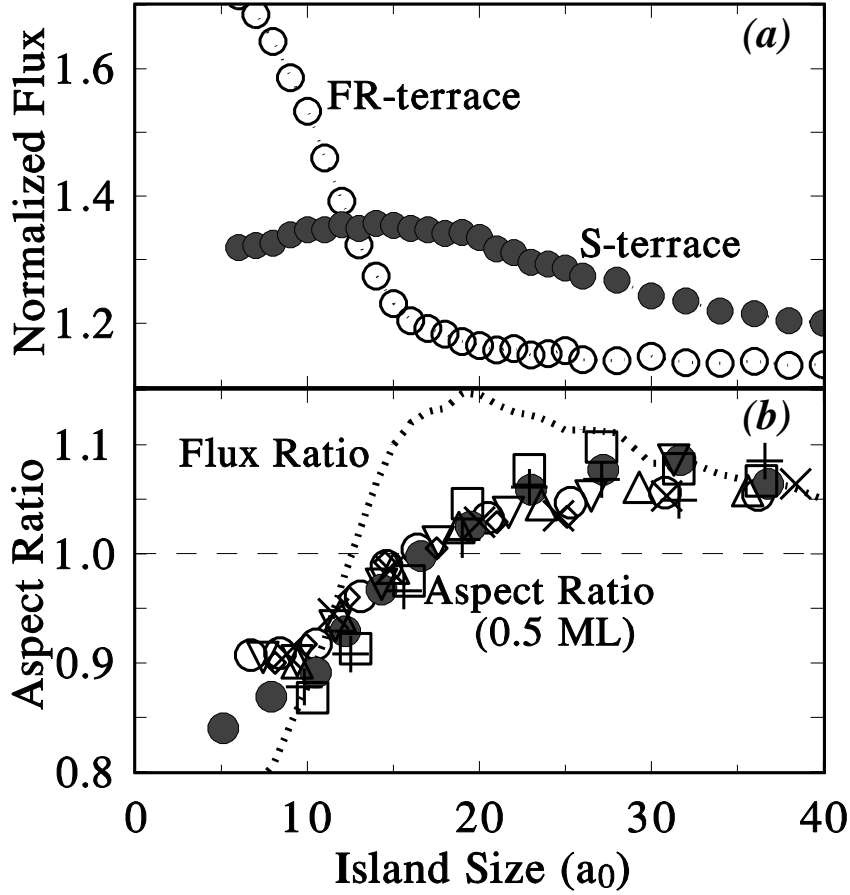}
 \caption{\label{fig4} Deposition flux and the aspect ratio according
 to the island size. (a)The filled and open circles represent the
 average deposition flux on $S$-terrace($F_{\rm S-terrace}$) and
 that on $FR$-terrace($F_{\rm FR-terrace}$), respectively.
 Each flux is normalized to the mean flux.
 (b)The ratio of deposition fluxes,
 ($F_{\rm S-terrace}/F_{\rm FR-terrace}$; dotted curve)
 and the aspect ratio ($W/L$) for 0.5 ML film grown with
 various temperatures (190 K-300 K) and diffusion parameters
 (open circle(Table I),
 +($\Delta E_{Sch}=$0.05 eV),
 box($\Delta E_{Sch}=$0.00 eV),
 diamond($\Delta E_{Sch}=$0.15 eV),
 closed circle($E_{2}=$0.553 eV, $E_{4}=$0.485 eV),
 $\Delta$(deposition rate 1/2 $F_0$ ),
 X(deposition rate 1/12 $F_0$ ),
 $\nabla$ ($\nu_{0}=$ 1.2 $\times 10^{13}$ ))
 are plotted as a function of the island size.}
\end{figure}

\begin{figure}
\includegraphics[scale=0.8]{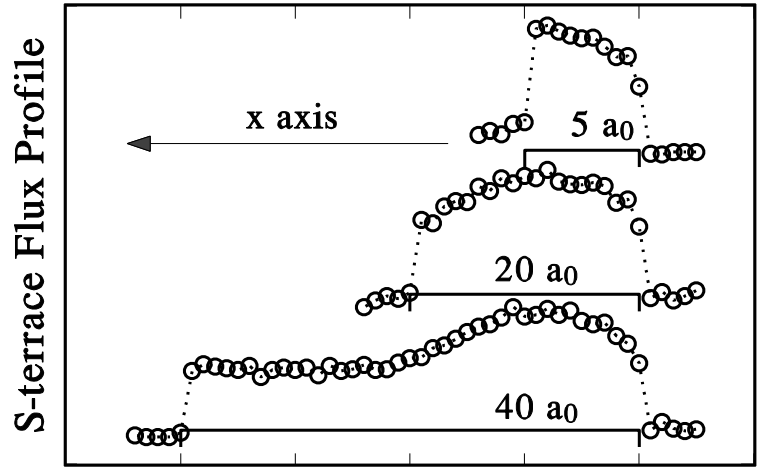}
\caption{\label{fig5} Flux profile on S-terrace along the side edge for
islands of varying sizes, 5, 20, and 40 $a_{0}$ long.
Solid lines represent both the position of pre-existing islands and
mean flux.}
\end{figure}

\end{document}